\documentclass[%
 reprint,
 amsmath,amssymb,
 longbibliography,
 aps,
 pre,
]{revtex4-1}

\usepackage[normalem]{ulem}
\usepackage{graphicx}
\usepackage{bm}
\usepackage{xcolor}
\usepackage{amsmath}
\usepackage[utf8]{inputenc}
\usepackage{mathtools}
\usepackage{xr}
\usepackage{caption}
\usepackage{subcaption}
\usepackage{svg}
\usepackage{soul}
\usepackage{bbm}
\usepackage{array}
\usepackage{hyperref}

\begin{document}

\title{Slow Dynamics and the Geometry of Jammed Packings}
% \title{Jammed Packings Geometry and Slow Dynamics}
\author{Eddie Bautista}
\email{ebautis2@uoregon.edu}

\author{Eric I. Corwin}
\email{ecorwin@uoregon.edu}
\affiliation{Department of Physics, Materials Science Institute, and Institute of Fundamental Studies, University of Oregon, Eugene, Oregon 97403, USA.}
\date{\today}

\begin{abstract}

Saddle points in the energy landscape of granular packings dominate the discrete steepest descent dynamics and ultimately determine the path that an out of mechanical equilibrium packing will follow and the resulting stable minimum that it will find. The saddle points that ultimately determine the resulting minima tend to be low-index saddle points. For models with an analytic energy landscape, such as the $p$-spin model, the steepest descent minimization path is affected by higher-index saddle points, which pull the system towards saddle points of decreasing index before arriving at the minima. Here, we examine the steepest descent minimization path of granular packings and compare them to the $p$-spin model. We show that the granular packing steepest descent minimization paths act like their smooth energy landscape counterparts and get attracted by saddle points. The index versus time curves for all models follow a shifted, stretched exponential. We further show that the shape parameter for the granular packings is unchanged when the energy landscape is modified to become analytic (Gaussian potential in a harmonic well) or non-local (Mari-Krzakala-Kurchan). The $p$-spin, on the other hand, has a significantly larger shape parameter. The reason is not due to the dimensionality, packing fraction, nonanalyticity, or the locality of the Hamiltonian of the models. The exact reason for the discrepancy in the shape parameter is \st{still} an unsolved mystery. 

\end{abstract}

\maketitle

%\section{Introduction}

\section{Introduction} Steepest descent energy minimization is both simple and straightforward \cite{cauchy_methode_nodate,suryadevara_basins_2025}. However, when dealing with high-dimensional spaces, one discovers that steepest descent acts very differently from the low-dimensional spaces upon which we build our intuition. A ball placed on a smooth and continuous $2d$ surface, for example, will roll straight down to a minima. This will seldom be true for high-dimensional spaces in which one observes extremely tortuous minimization paths \cite{hwang_understanding_2016,thirumalaiswamy_exploring_2022}.

Glasses are disordered systems which occupy a high-dimensional phase space and are characterized by extremely slow dynamics~\cite{castellani_spin-glass_2005}. Understanding the source of these slow dynamics is an open question ~\cite{gardner_spin_1985, sibani_anomalous_1987, fisher_equilibrium_1988, bouchaud_weak_1992,  kurchan_phase_1996}. Most proposed mechanisms rely on the system needing to overcome some type of energy barrier, thus requiring some type of activation energy in order for the systems to continue evolving~\cite{fisher_equilibrium_1988,bouchaud_weak_1992, sibani_anomalous_1987, gardner_spin_1985}. However, for spin glasses, a non-Arrhenius mechanism for slow dynamics has been proposed that depends solely on the geometry of phase space as tessellated into basins of attraction~\cite{kurchan_phase_1996}. A basin of attraction is the set of all configurations that minimize to the same state under perfectly overdamped continuous dynamics~\cite{ashwin_calculations_2012, martiniani_when_2023}. In this picture, saddle points in the energy landscape attract the system as it loses energy, thereby slowing the dynamics by pulling it towards nearby saddle points before it reaches the energy minima. Since the path is always near a saddle point, the index of the nearest saddle point in phase space is well approximated by the number of negative eigenvalues of the Hessian at each point along its path. This argument, however, relies on 1) a smooth energy landscape with no infinite energy density configurations and 2) infinite phase-space dimensions \cite{kurchan_phase_1996}. 

Amorphous granular packings offer a zero temperature analogue of glasses with no infinite energy density configurations \cite{gardner_spin_1985, charbonneau_jamming_2015, charbonneau_numerical_2015, dennis_jamming_2020, artiaco_exploratory_2020}, and saddle points in their energy landscape have been shown to dominate their dynamics \cite{bautista_numerically_2025}. This gives credence to the idea that the same non-Arrhenius mechanism in spin glasses is at play in granular packings. However, unlike spin glasses, granular packings do not have a smooth energy landscape, as they interact through a non-analytic contact potential \cite{sciortino_potential_2005, speedy_random_1998, martiniani_numerical_2017}. Additionally, they are exactly mean-field only in the limit of infinite spatial dimensions \cite{castellani_spin-glass_2005, sartor_mean-field_2021}. In this letter, we numerically explore the dynamics of granular packing to answer the question of whether saddle points in the energy landscape slow down their dynamics in the same manner as described in reference \cite{kurchan_phase_1996}. We examine the steepest descent minimization path for random initial conditions for granular packings. At each minimization time step, we calculate the index of the nearest saddle point and propose that a shifted, stretched exponential is the best-fit function for the index versus time curve. To compare to the spin glass prediction in \cite{kurchan_phase_1996}, we repeat our methods with a spherical $p$-spin model and find that it too follows a shifted, stretched exponential function, albeit with a different shape parameter, $k$. 

In order to understand the discrepancy in shape parameters between the two glassy models, we look at the Gaussian potential in a harmonic well (GPHW) and the Mari-Krzakala-Kurchan (MKK) model. The GPHW tests whether the non-analyticity of the granular packing is to blame for the change in shape parameter, and the MKK tests whether the difference lies in the locality of the granular packing potential. Neither model recovers the $p$-spin shape parameter. We leave the reason why the shape parameter is different between the granular packings and the $p$-spin as an open question. 

\section{Models and Methods}
%%%%%%%%%%%%%%%%%%%%%%%%%%%%%%%%%%%%%%%%%
\subsection{Granular Packing} 

\paragraph*{Model ---} We define our granular packing (GP) model to be consistent with \cite{bautista_numerically_2026} as a set of $N$ harmonic soft spheres in a $d$-dimensional box of side length one with periodic boundary conditions. The Hamiltonian of the system with particle positions $\{\vec x_i\}$ and radii $\{r_i\}$ is written as
\begin{equation} \label{gp energy}
\mathcal{H}(\{\vec x_i, r_i\}) =\frac{1}{2} \sum_{ij} \left ( 1-\frac{ \rho_{ij}}{r_{ij}} \right ) ^2 \Theta \left( 1-\frac{\rho_{ij}}{r_{ij}} \right) ,
\end{equation}
where $\rho_{ij} = \| \vec{x}_i-\vec{x}_j\|$, $r_{ij}=r_i+r_j$, and $\Theta$ is the Heaviside function. The Heaviside function (which is necessary for contact potentials) makes granular systems non-analytic. Additionally, granular packings are known to be mean-field in the limit of infinite dimensions \cite{parisi_mean-field_2010, arceri_jamming_2022} but deviate from this limit in finite dimensions \cite{parisi_mean-field_2010, charbonneau_jamming_2015, sartor_mean-field_2021}. 

Each energy landscape for this Hamiltonian is uniquely defined by the packing spatial dimension, $d$, number of particles, $N$, and set of particle radii, $\{r_i\}$. In $2d$, we choose particle radii from a log-normal distribution with a polydispersity (ratio of standard deviation to mean) of 0.25 to avoid spontaneous crystallization. In higher dimensions where crystallization is no longer a problem, the packings are monodispersed.

\paragraph*{Dynamics ---} Differentiating the Hamiltonian and taking the negative gives us the $\alpha$ component of the force on the $i$th particle,
\begin{equation} \label{gp force}
F_i^\alpha(\{\vec x_i, r_i\}) = \sum_{k}{\left(1-\frac{ \rho_{ij}}{r_{ij}} \right) \frac{n_{ik}^\alpha}{r_{ij}}},
\end{equation}
where $n^\alpha_{ik}$ is the $\alpha$-component of the normal vector pointing from particle $i$ to particle $k$. Using this force equation, we define our dynamics as, 
\begin{equation} \label{gp dynamics}
    \frac{\partial \{\vec x_i^\alpha\}}{\partial t} = F_i^\alpha(\{\vec x_i, r_i\}).
\end{equation}
We discretize these dynamics to form a steepest descent (SD) minimizer with a fixed time step, $\delta$, which we write as
\begin{equation} \label{gp SD}
\{ \vec{x}_{i,t+1}^{\alpha} \} = \{ \vec{x}_{i,t}^{\alpha} \} + \delta F_i^\alpha(\{\vec x_i, r_i\}),
\end{equation}

Taking the second derivative of the Hamiltonian gives the Hessian as,
\begin{equation} \label{gp hessian}
\begin{split}
H_{ij}^{\alpha\beta}\left(\{\vec x_i, r_i\}\right) &= \delta_{ij} \sum_{k}\left( \frac{1}{r_{ij}^2}n^{\alpha}_{ik}n^{\beta}_{ik} + \frac{\varepsilon_{ik}}{\rho_{ik}r_{ij}}(n^{\alpha}_{ik}n^{\beta}_{ik}-\delta^{\alpha\beta})  \right) \\
    & \quad - \delta_{\langle ij \rangle}\left( \frac{1}{r_{ij}^2}n^{\alpha}_{ij}n^{\beta}_{ij} + \frac{\varepsilon_{ij}}{\rho_{ij}r_{ij}}(n^{\alpha}_{ij}n^{\beta}_{ij}-\delta^{\alpha\beta}) \right)
\end{split}
\end{equation}
where $\varepsilon_{ij}=1-\frac{ \rho_{ij}}{r_{ij}}$ (the dimensionless overlap), $\delta_{ij}$ is the dirac-delta function and $\delta_{\langle ij \rangle}$ denotes particles in contact.

\paragraph*{Measurement ---} We explore the dynamics by sampling 100 random initial states (i.e. infinite temperature configurations) uniformly across the landscape. We choose to start at random configurations so that we can employ a flat measure across the entire energy landscape. We do this by randomly placing the $N$ particles in a $d$-dimensional space using a Poisson distribution.  From the initial states, we use the steepest descent minimizer in Eq. \ref{gp SD} as implemented by the \textsc{pyCudaPacking} software ~\cite{morse_geometric_2014, charbonneau_universal_2016, morse_echoes_2017} to minimize the energy. \textsc{pyCudaPacking} is a GPU-based sphere packing minimizer. We use a time step of $\delta=5\times10^{-5}$ in natural units. We choose this time step because it provides smooth dynamics in a reasonable time frame \cite{bautista_numerically_2026}. At each SD time step, we calculate the Hessian of the configuration and diagonalize it to calculate the eigenvectors and eigenvalues. The index of the nearest saddle point is the number of negative, non-trivial eigenvalues. We average our data by first truncating the paths to the same size, then averaging over the 100 paths.

We sample the landscapes of $2d$, $3d$, $4d$, and $5d$ packings at or above the jamming packing fraction, $\varphi_J$. For a $2d$ packing with a polydispersity of 0.25 we use $\varphi_J=0.842$. The jamming packing fractions for monodispersed particles are $\varphi_J=0.6471$ in $3d$, $\varphi_J=0.4636$ in $4d$, and $\varphi_J=0.3155$ in $5d$ \cite{morse_geometric_2014}. For $d=2$ and 3, we explore $N=16$, 32, 64, and 128. In $4d$ we examine $N=32$, 64, and 128 and in $5d$ $N=70$ and 128. The lower $N$ limit is set to avoid self-interactions in periodic boundary conditions. 

%%%%%%%%%%%%%%%%%%%%%%%%%%%%%%%%%%%%%%%%%
\subsection{$p$-spin} 

\paragraph*{Model ---}The $p$-spin model is defined as in \cite{folena_rethinking_2020}. Specifically, we use a pure, spherical $p$-spin model with $p=3$. The Hamiltonian for such a model is defined as 
\begin{equation} \label{pSpin energy}
\mathcal{H}(\{\sigma_i\}) =-\sum_{i<j<k}^{N} J_{ijk}\sigma_i\sigma_j\sigma_k,
\end{equation}
where $J_{ijk}$ is a rank 3 tensor of coupling constants drawn as independent variables from a Gaussian distribution with mean zero and variance $p!/2N^{p-1}$, $N$ is the number of particles, and $\sigma_i$ is the spin of particle i~\cite{folena_rethinking_2020}. We note that the $p$-spin model is exactly mean-field, has an analytic energy landscape, and the system does not have a sense of locality or sparseness because every particle interacts with every particle \cite{castellani_spin-glass_2005, derrida_random-energy_1980, derrida_random-energy_1981, crisanti_sphericalp-spin_1992}. The spins, $\sigma_i$, are real variables subject to the global spherical constraint
\begin{equation} \label{pSpin constraint}
\sum_i\sigma_i^2=N.
\end{equation}

\paragraph*{Dynamics ---}In the same vein as the granular packings, we define the dynamics of the $p$-spin model with spins $\{\sigma_i\}$ as
\begin{equation} \label{pSping dynamics}
\frac{\partial \{\sigma_i\}}{\partial t} = F_i(\{\sigma_i\}) =-\mu\left(\{\sigma_{i}\}\right)*\{\sigma_i\} - \frac{\partial \mathcal{H}}{\partial \{\sigma_i\}},
\end{equation}
where $\mu\left(\{ \sigma_{i}\}\right)=-\frac{1}{N}\sum_i\sigma_i\frac{\partial H}{\partial\sigma_i}$ is a Lagrange multiplier used to enforce the constraint in Eq. \ref{pSpin constraint}. We discretize these dynamics to form a steepest descent minimizer (SD) for the $p$-spin model as,
\begin{equation} \label{pSping SD}
\{\sigma_{i, t+1}\} = \{\sigma_{i, t}\} - \delta \left (\mu\left(\{\sigma_{i, t}\}\right)  + \frac{\partial \mathcal{H}_{i,t}}{\partial \{\sigma_{i,t}\}} \right),
\end{equation}
where $\{\sigma_{i,t}\}$ are the spins at time t. Similarly, we hold $\delta$ constant throughout the minimization process. 

The Hessian of the spherical $p$-spin is 
\begin{equation} \label{pSping hessian}
H_{ij}\left( \{ \sigma_i\}\right) = \frac{\partial^2 \mathcal{H}}{\partial \{ \sigma_i\} \partial \{\sigma_j\}}  -\mu\{\sigma_{i, t}\}*\delta_{ij},
\end{equation}
where $\delta_{ij}$ is the dirac-delta function.

\paragraph*{Measurement ---}We explore the dynamics by sampling 100 random initial states uniformly across the landscape, which can be thought of as the infinite-temperature limit of the system. We do this by choosing a random point on the surface of the $N$-dimensional sphere to satisfy the constraint in Eq. \eqref{pSpin constraint}. 

From the initial states, we use the steepest descent minimizer in Eq. \ref{gp SD} to minimize the energy. At each SD time step, we calculate the Hessian of the configuration and diagonalize it to calculate the eigenvectors and eigenvalues. The index of the nearest saddle point is the number of negative, non-trivial eigenvalues. As before, we average our data by first truncating the paths to the same size, then averaging over the 100 paths.

We sample the landscapes of $N=20$, 500, 1000. $N=20$ is meant to be a direct comparison to the granular packing case. As such, we use the same time step of $\delta=5\times10^{-5}$. This time step is too small to minimize the system in a reasonable amount of time in the large $N$ limit. Therefore, for $N\ge500$, we increase $\delta$ to $1\times10^{-2}$.

\section{Results} 
In section \ref{subsec:index}, we show that the index for the granular packings decreases monotonically as a function of time under steepest descent dynamics and are well fit by a shifted, stretched exponential function. We compare to the $p$-spin and find they share the same functional form, but with different shape parameters, i.e., stretching exponent. In section \ref{subsec:force}, we plot the force magnitude projected onto the positive and negative curvature subspaces. We show that the force primarily points in the positive curvature direction, thus indicating that the path sticks to the border between the basins of attraction. In section \ref{subsec:shapeParameter}, we discuss why the granular packings and the $p$-spins have different shape parameters. Finally, in section \ref{subsec:startIndex}, we discuss why the starting index for granular packings is not $Nd/2$, as may be naively expected by starting at a random starting configuration. The deviation from $Nd/2$ is due to an asymmetry between high and low energies in the energy landscape of granular packings. 

\subsection{Index Versus Time} \label{subsec:index}
\begin{figure}[t!]
\centering
\includegraphics[width=\columnwidth]{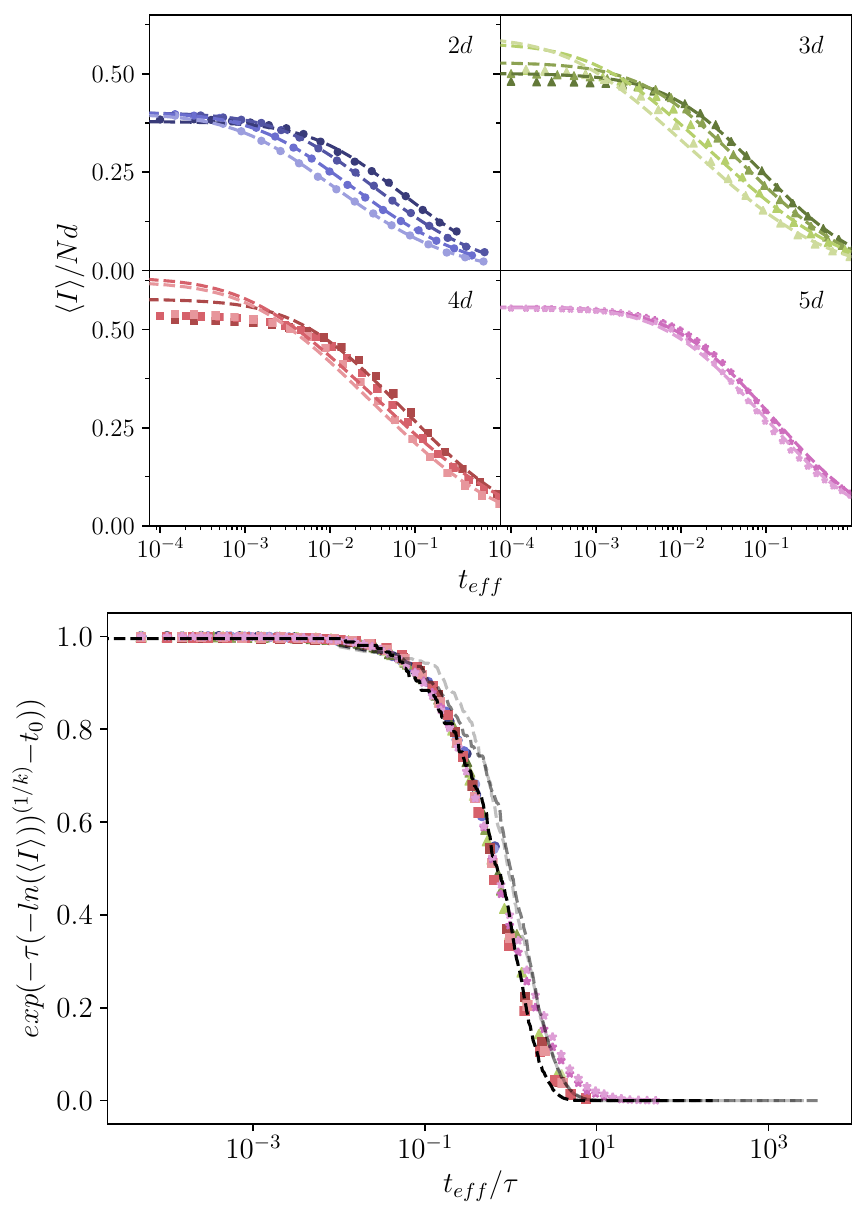}
\caption{
(Top) Steepest descent paths of index versus time for granular packings in spatial dimension $d=2$ (circles), 3 (triangles), 4 (squares), and 5 (stars) spatial dimensions, $d$, at a constant packing fraction of $1.07\varphi_J$. Particle number, $N$, is sampled from a range between 16 and 128 particles, with lighter colors representing larger $N$. The y-axis shows the index scaled by the total dimensionality of the energy landscape, $Nd$, averaged over 100 paths. The x-axis shows the time elapsed. The dashed lines are fits to a shifted, stretched exponential function, Eq.~\eqref{ssExp}. (Bottom) Scaled average index vs time paths for all data in the top figure. The time scale, $\tau$, time shift, $t_0$, and shape parameter, $k$, are scaled out from the y-axis. The black/gray dashed line shows the results for a $p$-spin model with $p=3$ and $N=20,$ 500, and 1000. Lighter black/gray lines represent larger $N$.}
\label{AvgI}
\end{figure}

Fig. \ref{AvgI} (top) shows the average index, $I$, scaled by the total number of degrees of freedom versus time plots for granular packings.  We see the index decrease monotonically over time, as predicted by \cite{kurchan_phase_1996}. The decrease in index is well fit by a shifted, stretched exponential function of the form
\begin{equation} \label{ssExp}
I(t) = e^{-\left(\frac{t+t_0}{\tau}\right)^k}.
\end{equation}

The scale parameter, $\tau$, determines how quickly the saddle point index decreases with time. Here, $\tau$ has a slight decrease with particle number and a slight increase with spatial dimension, as seen in Table \ref{tab:fitparams}. 

The shape parameter, $k$, determines the change in the rate at which the saddle index decreases and characterizes the amount of stretch imposed on an exponential function. For all systems, $k < 1$, meaning that the rate decreases with time. At the beginning of the minimization, the saddle index decreases at a faster rate than towards the end. $k$ is seemingly independent of $N$ and $d$ as seen in Table \ref{tab:fitparams}.

The time shift, $t_0$, moves the origin of the stretched exponential to negative times. This tells us that we are not starting at a maximum in the energy landscape. Instead, our initial configuration is somewhere along the minimization path and at an intermediate index. Increasing the dimensionality of the energy landscape should place the random starting configuration closer to the boundaries between basins of attraction and, consequently, increase the starting energy and lower $t_0$ (though not necessarily setting it to zero, since we are not guaranteed to start at a maximum). This is, in fact, what we see in Table \ref{tab:fitparams}. Increasing $d$ slightly lowers $t_0$ and increasing $N$, strongly decreases $t_0$. 

The bottom figure in Fig. \ref{AvgI} shows all data collapsed onto a single exponential master curve. To compare to the spin glass predictions in \cite{kurchan_phase_1996}, we add the collapsed curve for a $p$-spin with $N=20,$ 500, 1000 (dashed black/gray lines). The fit parameters are found in Table \ref{tab:fitparams}. Notably, the shape parameters for the $p$-spins are much greater than those for the granular packings, meaning that the structure of their energy landscapes is different. Nevertheless, the fact that granular packings share the same functional form as the $p$-spins provides evidence that the same non-Arrhenius mechanism for slow dynamics is at play for granular packings.

\subsection{Force Direction} \label{subsec:force}
The eigenvalues of the Hessian report on the curvature of the energy landscape, showing both the magnitude and sign of the curvature in each orthogonal direction. If a minimization path starts at or near the border between basins of attraction and were to strictly follow the positive curvature directions (positive eigenvalues), it would remain on the border and flow to subsequent saddle points. Conversely, if the path were to strictly follow the negative curvature directions (negative eigenvalues), then it would move away from the border and away from further saddle points. One would naively expect the minimization path to strictly follow the negative curvature direction because, in low dimensions, we associate negative curvature with the downhill direction.

\begin{figure}[t!]
\centering
\includegraphics[width=\columnwidth]{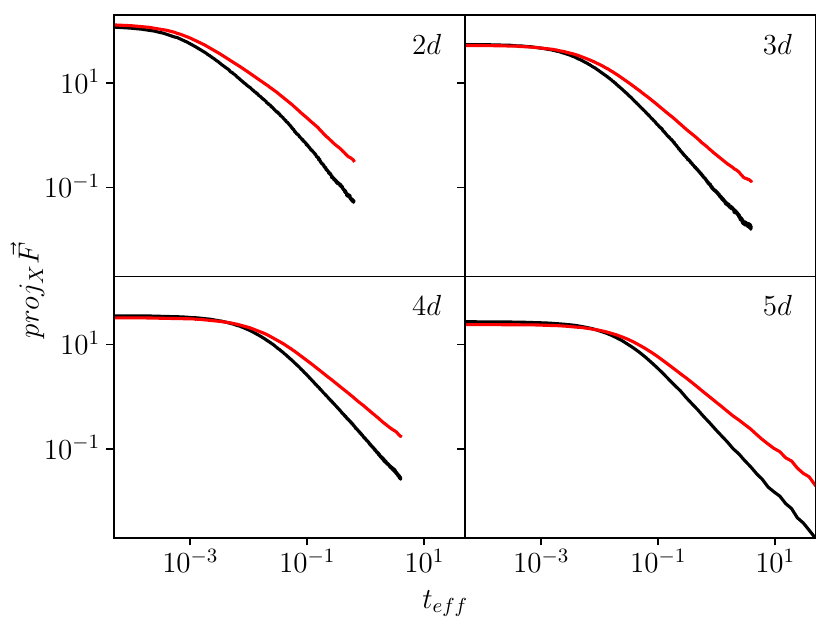}
\caption{
Force projection plot as a function of time for $d=2$, 3, 4, and 5 spatial dimensions and $N=128$. The red lines are the magnitude of the force projected onto the positive curvature subspace, and the black lines are the magnitude of the force projected onto the negative curvature subspace. The curves are averaged over 100 paths. 
}
\label{forcePlot}
\end{figure}

In Fig \ref{forcePlot}, we plot the magnitude of the force projected onto the positive (red) and negative (black) curvature subspaces as a function of time. The force primarily points in the positive curvature direction, which is the opposite of the naive expectation. It is important to note that this does not mean the energy is increasing. Instead, the path is minimizing its energy in the direction parallel to the border between basins of attraction to saddle points. This behavior, paired with the decreasing monotonic behavior between index and time, tells us that the path is following saddle points of decreasing index as described in \cite{kurchan_phase_1996}. 

\subsection{Shape Parameter} \label{subsec:shapeParameter}
We have shown evidence that the slow dynamics of granular packings is due to the saddle points in the energy landscape. However, the granular and the $p$-spin models exhibit different shape parameters with the shape parameter of granular packings being significantly smaller than that of the $p$-spin model. Here, we examine five possible reasons for this: 1) the systems we have examined are not at infinite phase space dimensions, 2) the packings are not mean-field, 3) the minimization paths of granular packings encounter a high density of discontuities in the landscape 4) the landscape is non-analytic and the minimization path frequently encounters discontinuities, and 5) these models differ in their sparseness or sense of locality (i.e. granular particles interact only with their neighbors while the $p$-spin interaction is all-to-all). We conclude that making the granular packings more mean-field and making the energy landscape smoother increases the shape parameter, but not enough to meet the $p$-spin shape parameter. 

\begin{figure}[t!]
\centering
\includegraphics[width=\columnwidth]{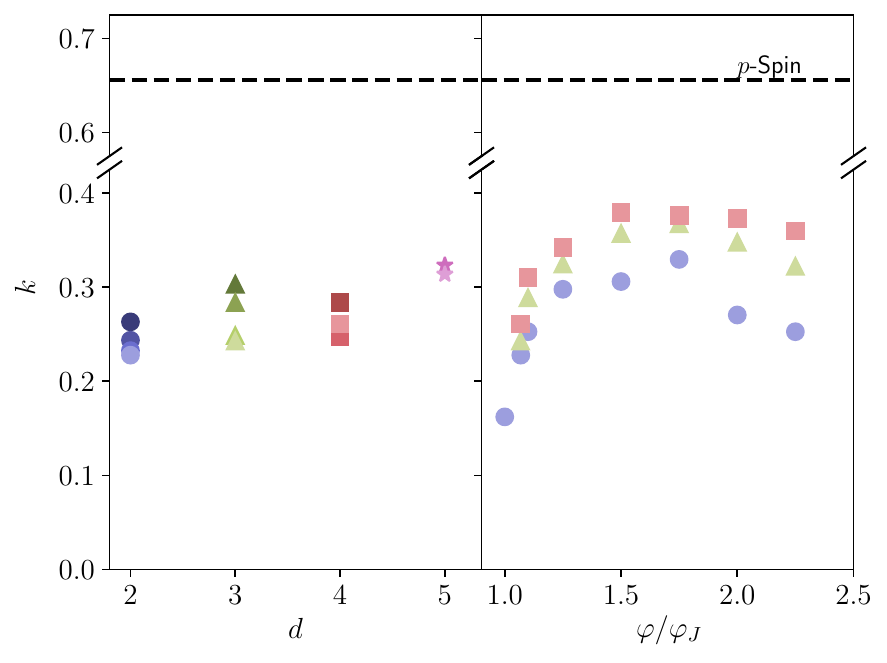}
\caption{
Plot of the shape parameter, k, as a function of spatial dimension (left) and packing fraction (right). For both plots, lighter colors represent larger particle numbers. Marker color and shape represent different spatial dimensions, with $2d$ data having blue circle markers, $3d$ green triangles, $4d$ red squares, and $5d$ pink stars. The left plot is held at a constant packing fraction of $1.07\varphi_J$. The dashed black line is the shape parameter of a $p$-spin model with $p=3$ and $N=20$.
}
\label{kplot}
\end{figure}

To examine 1) whether the finite phase space dimension is responsible for the different shape parameters, we plot the shape parameter for granular packings as a function of dimension in Fig. \ref{kplot} (left). The black dashed line shows the shape parameter for the $p$-spin. We see that increasing the phase space dimensions, $Nd$, does not significantly increase the $k$ parameter. The difference in k parameters between the lowest and highest phase space dimension is only 0.05. Since we observe only a very slow dependence in the granular shape parameter on dimension, it is unlikely that they obtain the same value as the $p$-spin model until extremely high spatial dimension, if at all.

It is unclear whether the dynamics of low-dimensional granular packings should be thought of as in the mean-field. It has been shown that important structural aspects of granular packings obey mean-field predictions all the way down to dimension $d=2$, such as bulk mechanical properties~\cite{sartor_mean-field_2021}, but dynamical properties are far less well studied.

The eigenvalues for a $p$-spin model follow a Wigner semicircle distribution that shifts towards positive eigenvalues as energy is minimized ~\cite{kurchan_phase_1996, auffinger_random_2010, fyodorov_hessian_2018, xu_hessian_2025}. The shift does not affect the shape of the distribution~\cite{kurchan_phase_1996, auffinger_random_2010, fyodorov_hessian_2018, xu_hessian_2025}. Therefore, if the packings are mean-field, then we expect their spectrum to have a constant shape irrespective of the depth within the energy landscape ~\cite{cavagna_stationary_1998, bray_statistics_2007}. 

\begin{figure*}[t!]
\centering
\includegraphics[width=\textwidth]{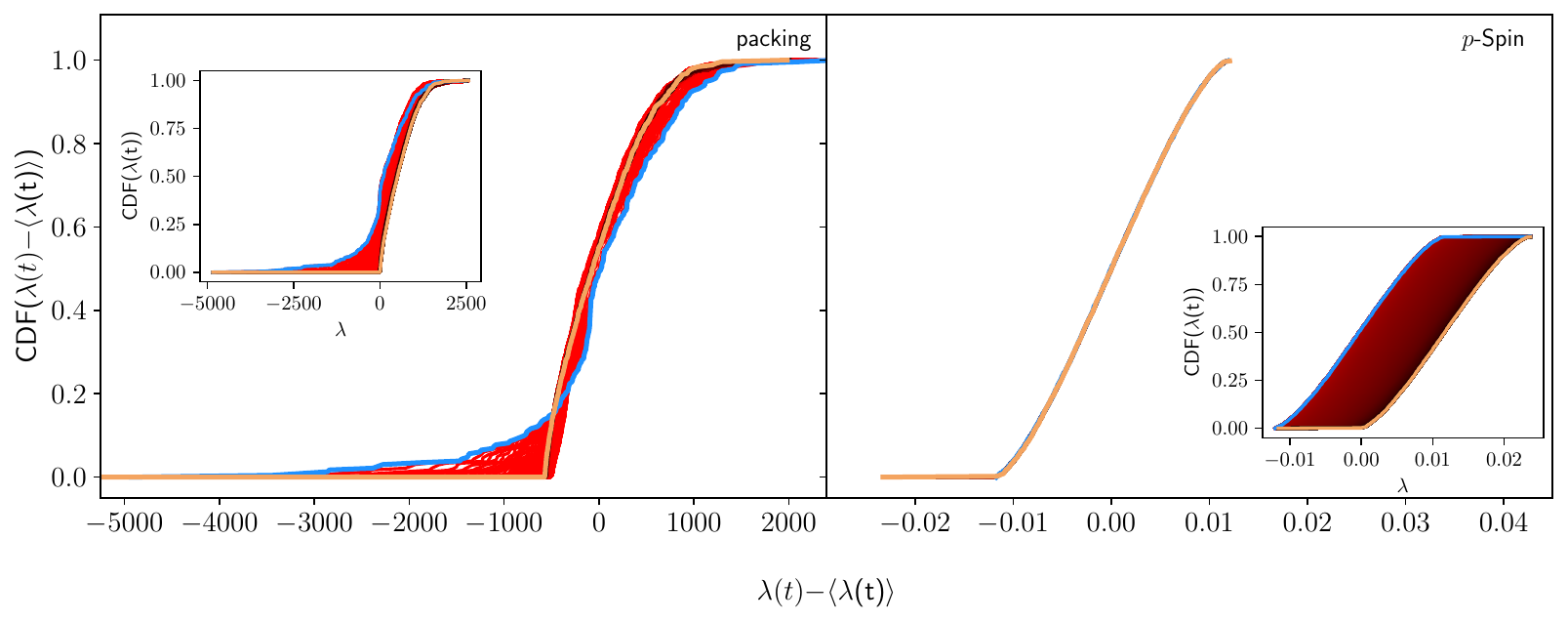}
\caption{
Cumulative distribution function (CDF) of eigenvalues at different times for granular packings (left) and $p$-spins (right). The plots are for a single minimization run from start to finish. Each curve shows the CDF at a different minimization time step, with darker colors representing later times. These curves show the evolving CDF along one minimization path for each model. The blue lines emphasize the initial configuration CDF, and the orange lines emphasize the last time step CDF. The main plots have the x-axis scaled by the average eigenvalue at each time step. The inset shows the unscaled CDFs. 
}
\label{spectrumPlot}
\end{figure*}

To examine whether 2) the granular packings being in the mean-fieldness of granular packings is responsible for the different shape parameters, we plot the time evolution of the cumulative distribution function (CDF) of the eigenvalues in Fig. \ref{spectrumPlot}. The left plot shows the CDFs for a $2d$ packing with $128N$. The right plot shows the CDFs for a $p$-spin system with $500N$ (right). Each curve represents a different time, from red at shortest times to black at longest times. The inset shows the CDF evolving with time, and the outset shows the collapsed CDFs achieved by subtracting off the mean eigenvalue at each time. We see the $p$-spin lines collapse completely onto the same curve, indicating that the shape does not change. The packing's CDFs, on the other hand, have a distribution that evolves significantly with time, with the distribution at the final time being narrower than the time zero curve. Thus, the evolution of the spectrum for the granular packing dynamics contributes additionally to the non-Arrhenius slow down.

\begin{figure}[t!]
\centering
\includegraphics[width=\columnwidth]{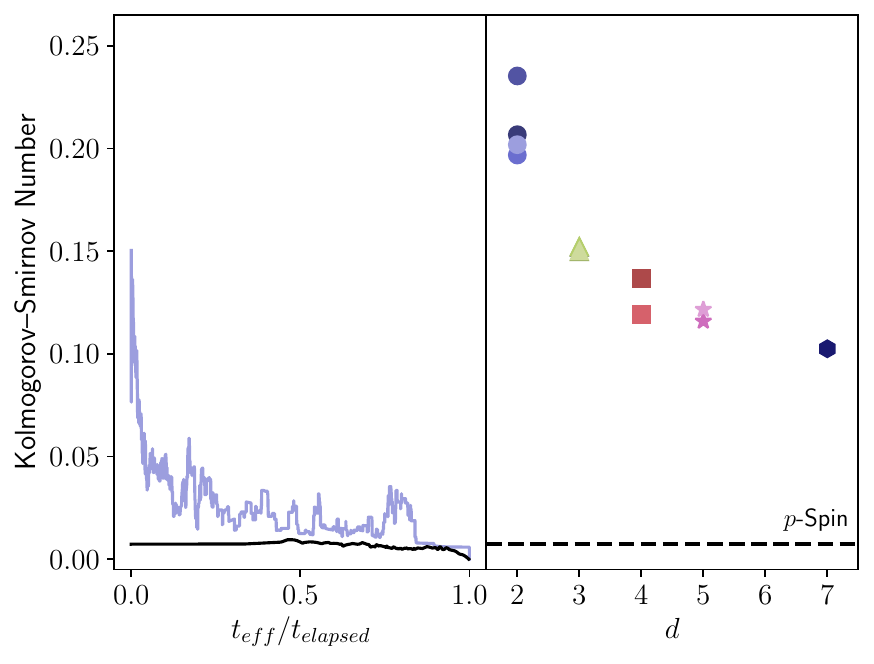}
\caption{
Plot of the average Kolmogorov-Smirnov (KS) number as a function of time (right) and spatial dimension (left). The right plot shows the KS number between the current time step and the final time step for the data shown in Fig. \ref{spectrumPlot}. The black curve shows the $p$-spin data. For the left plot, lighter colors represent larger particle numbers. The marker shapes and colors represent different spatial dimensions, with $2d$ data having blue circle markers, $3d$ green triangles, $4d$ red squares, $5d$ pink stars, and $7d$ purple hexagons. The black, dashed line shows the average Kolmogorov-Smirnov number for a $p$-spin model with $p=3$ and $N=500$.
}
\label{KS number}
\end{figure}

We characterize the difference between two CDFs by calculating the Kolmogorov-Smirnov (KS) number, i.e., the largest absolute distance between the two CDFs \cite{smirnov_table_1948, plato_n_2005}. In this case, the KS number characterizes how mean-field a system is, with a KS of zero being completely mean-field \cite{cavagna_stationary_1998, bray_statistics_2007}. Fig \ref{KS number} (left) shows the KS number between the CDF at time $t$ and the final CDF for the data in Fig \ref{spectrumPlot}. The blue curve is the granular packings and the black the $p$-spin. The plot shows that the change in shape of the distribution primarily happens at short times (bright red lines in Fig. \ref{spectrumPlot}) and quickly reaches the shape of the final distribution. The $p$-spin, on the other hand, remains near zero for the entire minimization, meaning that the CDF shape remains consistent throughout. 

Fig. \ref{KS number} (right) shows the average KS number between the initial CDF and the final CDF, averaged over all 100 paths, for all granular packings and (as a dashed line) the $N=500$ $p$-spin system. For packings, we see that as $d$ increases, the KS number decreases as expected. However, it is still far off from the KS number for the $p$-spins. Although the KS number for packings decreases with spatial dimension, it is still far from zero when $d=7$. 

At $\varphi_J$, the energy landscape of granular packings is known to be a fractally rough set of hierarchical sub-basins within sub-basins due to the infinite replica symmetry breaking \cite{charbonneau_glass_2017, artiaco_exploratory_2020, charbonneau_exact_2014, dennis_jamming_2020,charbonneau_fractal_2014,thirumalaiswamy_exploring_2022, rainone_following_2015,urbani_shear_2017}. As the packing fraction increases, the energy landscape becomes smoother in two ways. The first is that going away from jamming decreases the roughness of the energy landscape \cite{charbonneau_fractal_2014, castellani_spin-glass_2005}. The second is that increasing the packing fraction decreases the number of discontinuities along the minimization paths, i.e., the number of broken/formed contacts. At low packing fractions, but still above the jamming transition, the initial states have large volumes of empty space. Consequently, in the short-time regime, the particles have room to push away from one another and break/form contacts. As the packings flow into their minima, the particles fill in the empty space; thus, stabilizing the contact network. At large, but still physical, packing fractions, the volume of empty space decreases; thus, there are not as many contacts broken/formed in the short-time regime.

To examine whether 3) the discontinuities in the granular packing energy landscape are responsible for the different shape parameters, we plot the shape parameter for granular packings as a function of $\varphi$ in Fig. \ref{kplot} (right). We see $k$ change non-monotonically with packing fraction. The maximum $k$ achieved is about 0.4, which still falls far short of the $p$-spin shape parameter of 0.65. We note that the largest packing fractions searched here are not physical.   

\begin{figure}[t!]
\centering
\includegraphics[width=\columnwidth]{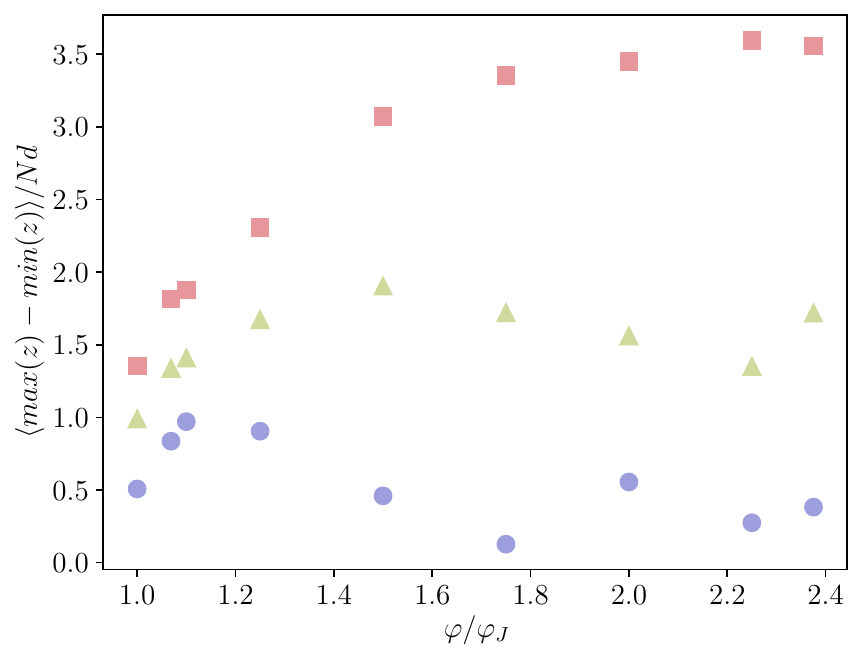}
\caption{
Average difference between the largest number and the smallest number of contacts along the minimization path as a function of packing fraction. All packings are held at 128 particles. Blue circle markers are $2d$ packings, green triangles are $3d$ packings, and red squares are $4d$ packings.
}
\label{contactDiff}
\end{figure}

To characterize the change in contacts during minimization, we plot the average difference between the maximum and the minimum number of contacts along the minimization path as a function of packing fraction in Fig. \ref{contactDiff}. In $2d$ and $3d$, we see similar non-monotonic behavior as in Fig. \ref{kplot} (right). In both cases, as the difference between the maximum and the minimum number of contacts decreases, the shape parameter increases. In $4d$, we do not see the same behavior, but it is likely hidden at larger packing fractions.

To examine whether 4) the non-analyticity of granular packings is responsible for the different shape parameters, we repeat our methods with the Gaussian potential in a harmonic well (GPHW) model, as fully described in Appendix \ref{app:GPHW}. The GPHW model is a completely analytic model sharing many features with granular packings. The index for a steepest descent path of the GPHW is also well fit by a shifted, stretched exponential and collapses onto the exponential master curve. The average index versus time plot is shown in Fig. \ref{gaussianIndex} and its collapse onto the exponential master curve in Fig. \ref{Master}. The fit yields a shape parameter of 0.25 as seen in Table \ref{tab:fitparams}. This is consistent with granular packings with the same spatial dimension and number of particles. This demonstrates that the nonanalyticity of the granular packing model is not responsible for the difference in shape parameter compared to the $p$-spin.

Finally, to examine whether 5) the locality of the granular packing is responsible for the different shape parameter, we repeat our methods with the Mari-Krzakala-Kurchann (MKK) model, as fully described in Appendix \ref{app:MKK}. The MKK model breaks locality by randomly shifting particle positions with quenched, Gaussian-distributed shifts with a standard deviation of $\sigma$. The index for a steepest descent path of the MKK is also well fit by a shifted, stretched exponential and collapses onto the exponential master curve. The average index versus time plot is shown in Fig. \ref{MKKIndex} and its collapse onto the exponential master curve in Fig. \ref{Master}. The shape parameter is weakly dependent on $\sigma$  and approaches $k=0.228$ as seen in Table \ref{tab:fitparams}. This is consistent with the granular packing. Therefore, the locality of the granular packing model is not responsible for the difference in shape parameter compared to the $p$-spin.

\subsection{Starting Index of Granular Packings} \label{subsec:startIndex}
In a generic landscape with $k$ degrees of freedom, one might expect that a random point will have an index of $k/2$, indicating an equal number of directions with positive and negative curvature. Indeed, this is precisely what is found for the $p$-spin system for which we observe a distribution peaked about $I=N/2$ for random configurations. However, the granular packing landscape encodes an asymmetry between high and low energy.

\begin{figure}[t!]
\centering
\includegraphics[width=\columnwidth]{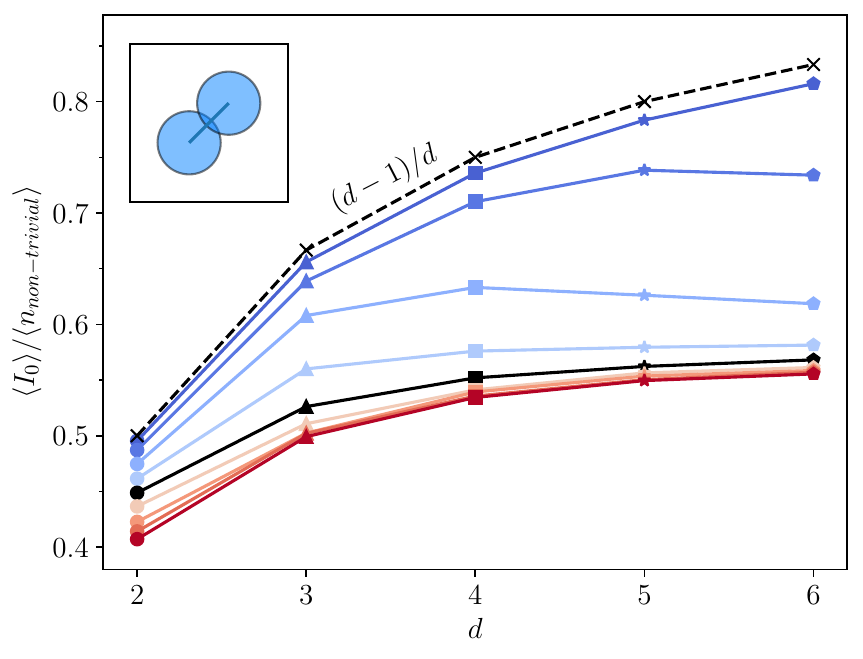}
\caption{
Plot of average starting index scaled by the number of non-trivial eigenvalues as a function of spatial dimension. The colors represent the packing fraction relative to the jamming packing fraction (black). Darker blues are further from jamming from below, and darker reds are further from above. The lowest packing fraction is $0.1\varphi$ and the largest $2.0\varphi$. The inset shows two overlapping particles.
}
\label{AvgStartingI}
\end{figure}

Fig. \ref{AvgStartingI} shows the average starting index scaled by the number of non-trivial eigenvalues vs dimension for different packing fractions. In the small packing fraction limit with random particle positions, we expect the packings to have a Poisson distribution of contact numbers, with most particles completely isolated, and most particles that are in contact to have only a single contact, as shown in Fig \ref{AvgStartingI} inset. Particles with only one contact will contribute $d$ non-trivial modes, of which one direction will have positive curvature in the energy landscape (in the direction of the point of contact), and $d-1$ directions will have negative curvature (the subspace perpendicular to the point of contact). Thus, one expects an index scaled by the number of non-trivial modes of $(d-1)/d$, as observed in Fig \ref{AvgStartingI} for the lowest density packings.

\section{Conclusion} These results show that the slow dynamics of granular packings arise from the same non-Arrhenius mechanism present in spin glasses  ~\cite{kurchan_phase_1996}, despite their energy landscape being neither smooth nor analytic. We demonstrate that the index versus time curves for granular packings and spherical $p$-spins decrease monotonically with time as a shifted stretched exponential. The time scale, $\tau$, has a slight decrease with particle number and a slight increase with spatial dimension. The time shift, $t_0$, decreases with spatial dimension and number of particles. The shape parameter, $k$, is different between the two models. 

We examine five possible reasons for this: 1) the systems we have examined are not at infinite phase space dimensions, 2) the packings are not mean-field, 3) the minimization paths of granular packings encounter a high density of discontinuities in the landscape, 4) the landscape is non-analytic, and 5) these models differ in their sparseness or sense of locality (i.e. granular particles interact only with their neighbors while the $p$-spin interaction is all-to-all). We conclude that making the granular packings more mean-field and making the energy landscape smoother increases the shape parameter, but not enough to meet the $p$-spin shape parameter. The reason why the models have different shape parameters despite both being glassy materials is left as an open question. 

We suspect the key to understanding the shape parameter lies in the structure of saddle points in the energy landscape. It might be the case that there is an asymmetry between the number of high- and low-index saddle points in the energy landscape. This is because, at a random starting configuration, the particles are often clustered together in highly unstable configurations, which can break up in many different ways, thus leading to a large number of high-index saddle points. At low energies, the packings are more stable, which restricts the movement of particles, thus leading to very few low-index saddle points. This mechanism is not present in the $p$-spin, thus leading to a more uniform distribution of the number of different index saddle points. 

%\section{Acknowledgments}

\section{Acknowledgments} We thank Jorge Kurchan and Peter Morse for useful discussions. This work was supported by a grant from the Simons Foundation No. 454939. The simulations were performed on the University of Oregon high-performance computing cluster, Talapas.

\appendix

\section{Gaussian potential in a harmonic well (GPHW)}
\label{app:GPHW}
%\subsection{Model} 
\paragraph*{Model ---}We designed our Gaussian potential in a harmonic well (GPHW) model to mimic the granular packing model, but with a completely analytic energy landscape. This would require its Hamiltonian to be everywhere represented by a convergent power series, contain the particles within a given volume, and exhibit a sense of locality (i.e., particles can only meaningfully interact with particles in their neighborhood). As such, we define the GPHW model as a set of $N$ particles trapped in a $d$-dimensional harmonic well, with each particle having a repulsive Gaussian potential dependent only on the distance between particles. The Gaussian potential is analytic everywhere, thus meeting the first criterion. The harmonic well meets both the first and second criteria. It pushes the particles towards its center, thus containing them within a given volume, and we do not have to worry about the energy landscape being disjointed at a periodic box boundary. The locality is achieved by an appropriate choice of the widths of the Gaussian potentials. Each particle is given a different width, $\sigma_i$, which we will call the particle radii. The radii are drawn from a log-normal distribution with a polydispersity of 0.25. These widths ensure that the particles only have strong interactions when in the same neighborhood and weak when far away. We define the Hamiltonian of the system with particle positions $\{\vec x_i\}$ and radii $\{\sigma_i\}$ as
\begin{equation} \label{gaussian energy}
\mathcal{H}(\{\vec x_i\}) = \sum_{ij} e^{\frac{- \rho_{ij}^2}{2\sigma_{ij}^2}} + \frac{k}{2}\sum_i\rho_{ij}^2,
\end{equation}
where, $\sigma_{ij} = \sqrt{\sigma_i^2 + \sigma_j^2}$ and $k=1$. 

%\subsection{Dynamics}
\paragraph*{Dynamics ---}The force can be written as,
\begin{equation} \label{gaussian force}
F_i(\{\vec x_i\}) = \sum_{k} \frac{(\vec{x}_i-\vec{x}_k)}{\sigma_{ik}^2}e^{\frac{- \rho_{ij}^2}{2\sigma_{ik}^2}} - k\vec{x}_i.
\end{equation}
The discretized dynamics are then
\begin{equation} \label{gaussian SD}
\{ \vec{x}_{i,t+1} \} = \{ \vec{x}_{i,t} \} + \delta F_i(\{\vec x_i\}).
\end{equation}
The Hessian of the Gaussian potential in a harmonic well is defined as,
\begin{equation} \label{gaussian Hessian}
\begin{split}
H_{ij}^{\alpha\beta} &= (1 - \delta_{ij}) e^{\frac{- \| \vec{x}_i-\vec{x}_j\|^2}{2\sigma_{ij}^2}} 
    \left( \frac{\delta_{\alpha\beta}}{\sigma_{ij}^2} - \frac{x_{ij}^\alpha x_{ij}^\beta}{\sigma_{ij}^4}\right) \\
    &\quad 
    \delta_{ij}\Biggl[\sum_l e^{\frac{- \| \vec{x}_i-\vec{x}_j\|^2}{2\sigma_{ij}^2}}
    \left( \frac{x_{ik}^\alpha x_{ik}^\beta}{\sigma_{il}^4} - \frac{\delta_{\alpha\beta}}{\sigma_{il}^2}\right) + k\delta_{\alpha\beta}\Biggr]
\end{split}
\end{equation}
where $x_{ij}^{\alpha} = (\vec{x}_i - \vec{x}_j)^\alpha$ and $\delta_{ij}$ is the dirac-delta function.

%\subsection{Measurement} 
\paragraph*{Measurements ---}We explore the dynamics by sampling 100 random initial states uniformly across the landscape, which can be thought of as the infinite-temperature limit of the system. The dynamics of the GPHW has two regimes. At short times, there is a short-lived compression regime due to the harmonic well pushing the particles towards the origin. After this initial compaction, there is a slow dynamic regime due to the particles rearranging. In this manuscript, we are not interested in the first regime. As such, we set the initial positions randomly and compactly near the origin to shorten the initial regime.

From the initial states, we use the steepest descent minimizer in Eq. \ref{gaussian SD} to minimize the energy. At each SD time step, we calculate the Hessian of the configuration and diagonalize it to calculate the eigenvectors and eigenvalues. The index of the nearest saddle point is the number of negative, non-trivial eigenvalues. As before, we average our data by first truncating the paths to the same size, then averaging over the 100 paths. In order to focus on the slow dynamic regime, we utilize a weighted fit on our data, with the tail of the curve being weighted more than the head.

We sample the landscapes of a $2d$ system with $N=128$. This system is meant to be a direct comparison to the granular packings case with the same spatial dimension and particle number. As such, we use the same time step of $\delta=5\times10^{-5}$.

%\subsection{Results}

\begin{figure}[t!]
\centering
\includegraphics[width=\columnwidth]{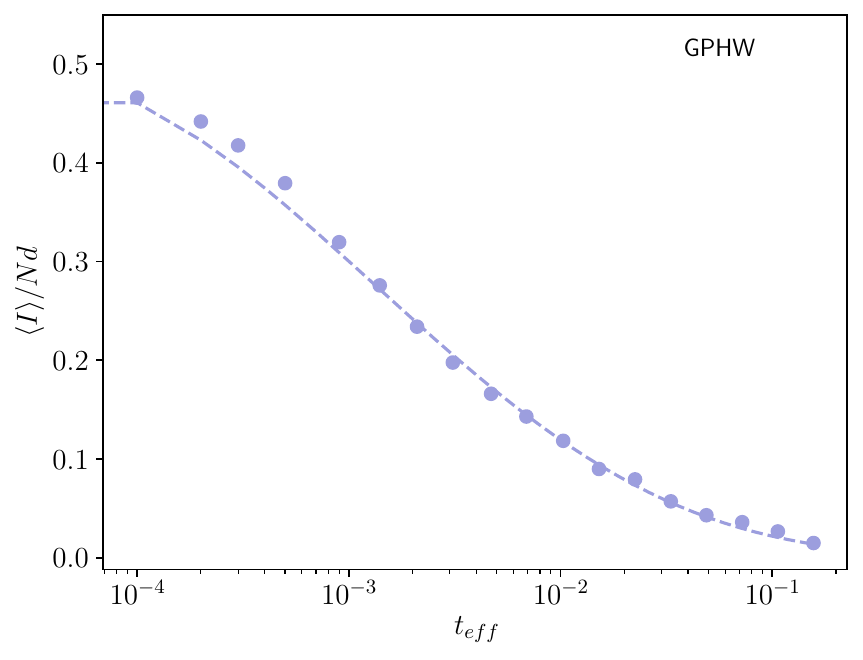}
\caption{
Steepest descent path of scaled index versus time for the Gaussian potential in a harmonic well (GPHW) for a system with $d=2$ and $N=128$. Markers represent the data and the dashed lines the fit to the shifted, stretched exponential function.
}
\label{gaussianIndex}
\end{figure}

\paragraph*{Results ---}Fig. \ref{gaussianIndex} shows the scaled index versus time plot for the GPHW model averaged over the 100 trials. The dashed line is the fit to the shifted, stretched exponential function. The index decreases monotonically with time and is well fit by a shifted stretched exponential function. 

The fit parameters are shown in Table \ref{tab:fitparams}. The shape parameter is consistent with that of the granular packings with the same spatial dimension, packing fraction, and number of particles. 

\section{Mari-Krzakala-Kurchann (MKK)} 
\label{app:MKK}
\paragraph*{Model ---}The Mari-Krzakala-Kurchann (MKK) model is a modification of the usual granular packing model by introducing a set of random pairwise shifts between particles \cite{mari_jamming_2009, mari_dynamical_2011}. These shifts allow particles that are not near to one another to interact, thereby breaking the locality typical of granular packing models. We write the shifted position of particle $j$ as seen by particle $i$ as, 
\begin{equation} \label{mkk shift}
x_j'^{\alpha} = x_j^{\alpha} + A_{ij}^\alpha
\end{equation}
where $A_{ij}^\alpha$ is a rank 3 tensor of quenched, Gaussian distributed random variables with a standard deviation of $\sigma$. $A_{ij}^\alpha$  is anti-symmetric in $ij$, meaning $A_{ij}^\alpha = -A_{ji}^\alpha$ \cite{nishikawa_collective_2022, charbonneau_jamming_2023}. By construction, the MKK model is mean-field in the limit of infinite $\sigma$ \cite{mari_jamming_2009}.

Since the only thing that changed is the particle locations, the functional forms of equations \eqref{gp energy} -- \eqref{gp hessian} remain the same except that every instance of $x_j$ or $x_k$ is replaced by $x_j'$ or $x_k'$. For example, \eqref{gp energy} becomes,
\begin{equation} \label{mkk energy}
\begin{split}
\mathcal{H}(\{\vec x_i\}) &= \frac{1}{2} \sum_{ij} \left ( 1-\frac{ \| \vec{x}_i-\vec{x}_j- A_{ij}^\alpha\|}{r_i+r_j+A_{ij}^\alpha} \right ) ^2 
    \\ & \quad \Theta \left( 1-\frac{ \| \vec{x}_i-\vec{x}_j-A_{ij}^\alpha\|}{r_i+r_j+A_{ij}^\alpha} \right).
\end{split}
\end{equation}
%

%\subsection{Measurement} 
\paragraph*{Measurements ---}We explore the dynamics by sampling 100 random initial states uniformly across the landscape, which can be thought of as the infinite-temperature limit of the system. We do this by randomly placing the $N$ particles in a $d$-dimensional space using a Poisson distribution.

From the initial states, we use the steepest descent minimizer in \eqref{gp SD}, with the shifted particle positions in Ep. \eqref{mkk shift}, as implemented by the \textsc{pyCudaPacking} software ~\cite{morse_geometric_2014, charbonneau_universal_2016, morse_echoes_2017} to minimize the energy. At each SD time step, we calculate the Hessian of the configuration and diagonalize it to calculate the eigenvectors and eigenvalues. The index of the nearest saddle point is the number of negative, non-trivial eigenvalues. We average our data by first truncating the paths to the same size, then averaging over the 100 paths.

We wish to sample the energy landscape of the MKK model as it approaches the mean-field limit in $2d$. Therefore, we simulate systems with $\sigma=0.1$, 1, and 10. Here, $\sigma$ is scaled to the nearest-neighbor distance, i.e., $(N/V)^{-1/d}$, where $V$ is the box volume. To avoid self-interactions, we set $N=1000$. To minimize the system within a reasonable time frame, we use monodisperse radii with a time step of $\delta=1\times10^{-4}$ in natural units. 

%\subsection{Results}

\begin{figure}[t!]
\centering
\includegraphics[width=\columnwidth]{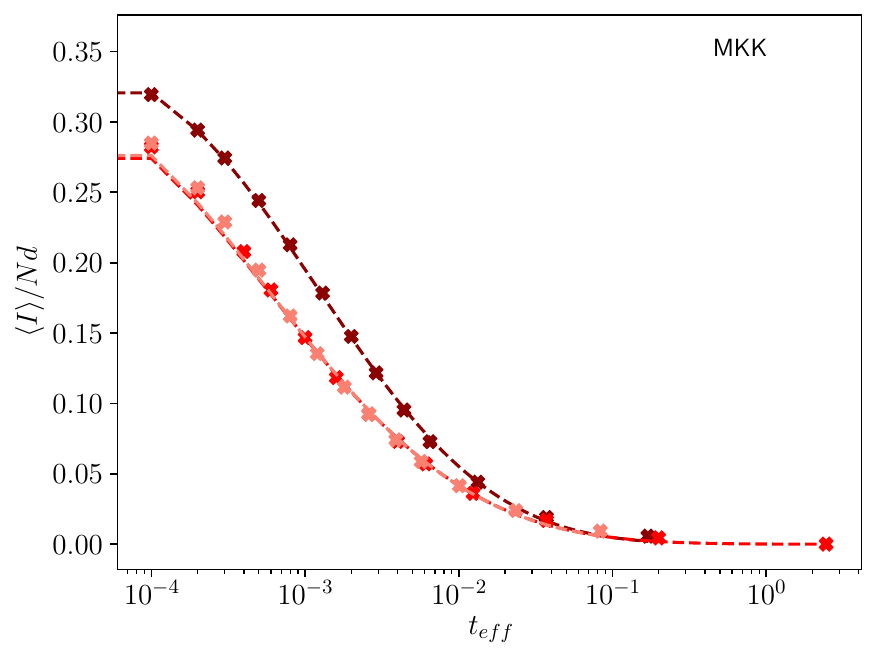}
\caption{
Steepest descent path of scaled index versus time for the MKK model. The MKK model has $d=2$ and $N=1000$. Darker colors in the MKK plot represent smaller $\sigma$ (standard deviation of the random shift tensor $A_{ij}^\alpha$). Markers represent the data, and the dashed lines represent the fit to the shifted, stretched exponential function.
}
\label{MKKIndex}
\end{figure}

\paragraph*{Results ---}Fig. \ref{MKKIndex} shows the scaled index versus time plot for the MKK model averaged over the 100 trials. The dashed line is the fit to the shifted, stretched exponential function. The index decreases monotonically with time and is well fit by a shifted stretched exponential function. As in the granular packing model, the average starting scales index is not $Nd/2$, showing the same asymmetry between high and low energy discussed in section \ref{subsec:startIndex}.

The fit parameters are shown in Table \ref{tab:fitparams}. The shape parameter at the mean-field limit (large $\sigma$) is $k\approx0.22$. At low $\sigma$, the shape parameter is only slightly higher, at $k=0.269$. 

\section{Master Curver}

\begin{figure}[t!]
\centering
\includegraphics[width=\columnwidth]{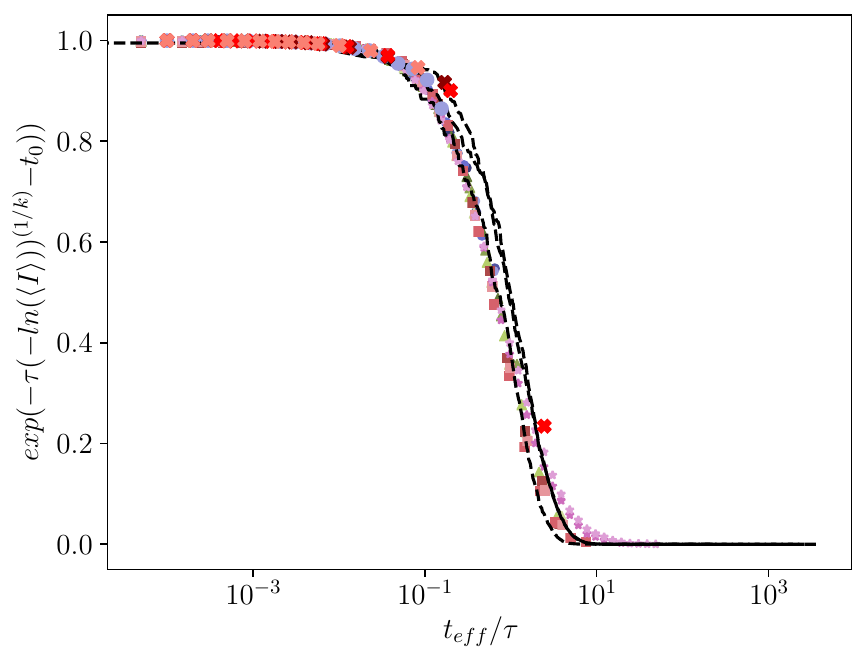}
\caption{
Scaled average index vs time paths for all data presented in this letter. The time scale, $\tau$, time shift, $t_0$, and shape parameter, $k$, are scaled out from the y-axis. The black dashed line shows the results for the $p$-spin models.
}
\label{Master}
\end{figure}

Fig. \ref{Master} shows the collapsed data for all models presented in the letter. We see all the data neatly collapse onto a single master curve.  

\section{Fit Parameters}
Fit parameters for all data presented in this letter.

% \setlength{\extrarowheight}{1.75pt}

% \begin{longtable}{|c|c|c|c|}
\begin{table}[h]
    \centering
    \caption{Shifted, stretched exponential fit parameters for granular packings (GP), $p$-spin, Gaussian potential in a harmonic well (GPHW), and Mari-Krzakala-Kurchann (MKK) systems.}
    \label{tab:fitparams}%\\ \hline
    \renewcommand{\arraystretch}{1.23}
    \begin{tabular}{|c|c|c|c|}  \hline
        model & $\tau$ & $k$ & $t_0$ \\ \hline
        GP: $2d$, $16N$, $1.07\varphi/\varphi_J$ & $1.25 \times 10^{-2}$  &  0.26  &  $1.12 \times 10^{-2}$ \\ \hline
        GP: $2d$, $32N$, $1.07\varphi/\varphi_J$ & $6.09 \times 10^{-3}$  &  0.24  &  $4.17 \times 10^{-3}$ \\ \hline
        GP: $2d$, $64N$, $1.07\varphi/\varphi_J$ & $3.04 \times 10^{-3}$  &  0.23  &  $2.01 \times 10^{-3}$ \\ \hline
        GP: $2d$, $128N$, $ 1.0\varphi/\varphi_J $ & $1.03 \times 10^{-03}$  &  0.16  &  $4.54 \times 10^{-04}$ \\ \hline
        GP: $2d$, $128N$, $1.07\varphi/\varphi_J$ & $1.79 \times 10^{-3}$  &  0.22  &  $1.23 \times 10^{-3}$ \\ \hline
        GP: $2d$, $128N$, $ 1.1\varphi/\varphi_J $ & $2.17 \times 10^{-03}$  &  0.25  &  $1.60 \times 10^{-03}$ \\ \hline
        GP: $2d$, $128N$, $ 1.25\varphi/\varphi_J $ & $1.91 \times 10^{-03}$  &  0.3  &  $1.37 \times 10^{-03}$ \\ \hline
        GP: $2d$, $128N$, $ 1.5\varphi/\varphi_J $ & $1.48 \times 10^{-03}$  &  0.31  &  $9.03 \times 10^{-04}$ \\ \hline
        GP: $2d$, $128N$, $ 1.75\varphi/\varphi_J $ & $1.69 \times 10^{-03}$  &  0.33  &  $1.19 \times 10^{-03}$ \\ \hline
        GP: $2d$, $128N$, $ 2.0\varphi/\varphi_J $ & $1.07 \times 10^{-03}$  &  0.27  &  $5.53 \times 10^{-04}$ \\ \hline
        GP: $2d$, $128N$, $ 2.25\varphi/\varphi_J $ & $1.01 \times 10^{-03}$  &  0.25  &  $5.20 \times 10^{-04}$ \\ \hline
        GP: $3d$, $16N$, $1.07\varphi/\varphi_J$ & $3.40 \times 10^{-2}$  &  0.30  &  $1.00 \times 10^{-2}$  \\ \hline
        GP: $3d$, $32N$, $1.07\varphi/\varphi_J$ & $2.22 \times 10^{-2}$  &  0.28  &  $4.52 \times 10^{-3}$ \\ \hline
        GP: $3d$, $64N$, $1.07\varphi/\varphi_J$ & $1.13 \times 10^{-2}$  &  0.24  &  $9.95 \times 10^{-4}$ \\ \hline
        GP: $3d$, $128N$, $1.07\varphi/\varphi_J$ & $6.91 \times 10^{-3}$  &  0.24  &  $4.60 \times 10^{-4}$ \\ \hline
        GP: $3d$, $128N$, $ 1.1\varphi/\varphi_J $ & 1.1 $1.02 \times 10^{-02}$  &  0.29  &  $2.16 \times 10^{-03}$ \\ \hline
        GP: $3d$, $128N$, $ 1.25\varphi/\varphi_J $ & 1.25 $1.01 \times 10^{-02}$  &  0.32  &  $2.70 \times 10^{-03}$ \\ \hline
        GP: $3d$, $128N$, $ 1.5\varphi/\varphi_J $ & 1.5 $9.96 \times 10^{-03}$  &  0.36  &  $3.17 \times 10^{-03}$ \\ \hline
        GP: $3d$, $128N$, $ 1.75\varphi/\varphi_J $ & 1.75 $1.02 \times 10^{-02}$  &  0.37  &  $3.46 \times 10^{-03}$ \\ \hline
        GP: $3d$, $128N$, $ 2.0\varphi/\varphi_J $ & 2.0 $9.97 \times 10^{-03}$  &  0.35  &  $3.17 \times 10^{-03}$ \\ \hline
        GP: $3d$, $128N$, $ 2.25\varphi/\varphi_J $ & 2.25 $9.51 \times 10^{-03}$  &  0.32  &  $2.76 \times 10^{-03}$ \\ \hline
        GP: $4d$, $32N$, $1.07\varphi/\varphi_J$ & $3.86 \times 10^{-2}$  &  0.28  &  $4.68 \times 10^{-3}$ \\ \hline
        GP: $4d$, $64N$, $1.07\varphi/\varphi_J$ & $2.20 \times 10^{-2}$  &  0.24  &  $9.31 \times 10^{-4}$ \\ \hline
        GP: $4d$, $128N$, $1.07\varphi/\varphi_J$ & $1.84 \times 10^{-2}$  &  0.26  &  $1.06 \times 10^{-3}$ \\ \hline
        GP: $4d$, $128N$, $ 1.1\varphi/\varphi_J $ & $2.61 \times 10^{-02}$  &  0.31  &  $4.65 \times 10^{-03}$ \\ \hline
        GP: $4d$, $128N$, $ 1.25\varphi/\varphi_J $ & $2.51 \times 10^{-02}$  &  0.34  &  $5.59 \times 10^{-03}$ \\ \hline
        GP: $4d$, $128N$, $ 1.5\varphi/\varphi_J $ & $2.52 \times 10^{-02}$  &  0.38  &  $6.93 \times 10^{-03}$ \\ \hline
        GP: $4d$, $128N$, $ 1.75\varphi/\varphi_J $ & $2.36 \times 10^{-02}$  &  0.38  &  $6.32 \times 10^{-03}$ \\ \hline
        GP: $4d$, $128N$, $ 2.0\varphi/\varphi_J $ & $2.36 \times 10^{-02}$  &  0.37  &  $6.23 \times 10^{-03}$ \\ \hline
        GP: $4d$, $128N$, $ 2.25\varphi/\varphi_J $ & $2.25 \times 10^{-02}$  &  0.36  &  $5.67 \times 10^{-03}$ \\ \hline
        GP: $5d$, $70N$, $1.07\varphi/\varphi_J$ & $5.96 \times 10^{-2}$  &  0.32  &  $1.11 \times 10^{-2}$ \\ \hline
        GP: $5d$, $128N$, $1.07\varphi/\varphi_J$ & $4.62 \times 10^{-2}$  &  0.31  &  $8.08 \times 10^{-3}$ \\ \hline
        $p$-spin, $20N$, $1.07\varphi/\varphi_J$ & $5.68 \times 10^{0}$  &  0.55  &  $2.75 \times 10^{0}$ \\ \hline
        $p$-spin, $500N$, $1.07\varphi/\varphi_J$ & $1.03 \times 10^{2}$  &  0.63  &  $5.74 \times 10^{1}$ \\ \hline
        $p$-spin, $1000N$, $1.07\varphi/\varphi_J$ & $2.10 \times 10^{2}$  &  0.65  &  $1.21 \times 10^{2}$ \\ \hline
        GPHW: $2d$, $128N$, $1.07\varphi/\varphi_J$ & $5.33 \times 10^{-4}$  &  0.25  &  $9.69 \times 10^{-5}$ \\ \hline
        MKK: $2d$, $1000N$, $\sigma=0.1$ & $5.33 \times 10^{-4}$  &  0.26  &  $9.69 \times 10^{-5}$ \\ \hline
        MKK: $2d$, $1000N$, $\sigma=1.0$ & $1.96 \times 10^{-4}$  &  0.22  &  $2.16 \times 10^{-4}$ \\ \hline
        MKK: $2d$, $1000N$, $\sigma=10.0$ & $6.17 \times 10^{-5}$  &  0.22  &  $9.24 \times 10^{-5}$ \\ \hline
% \end{longtable}
    \end{tabular}
\end{table}

\bibliographystyle{unsrt}
\bibliography{references}

@article{bray_statistics_2007,
	title = {Statistics of {Critical} {Points} of {Gaussian} {Fields} on {Large}-{Dimensional} {Spaces}},
	volume = {98},
	url = {https://link.aps.org/doi/10.1103/PhysRevLett.98.150201},
	doi = {10.1103/PhysRevLett.98.150201},
	number = {15},
	urldate = {2026-08-27},
	journal = {Physical Review Letters},
	publisher = {American Physical Society},
	author = {Bray, Alan J. and Dean, David S.},
	month = apr,
	year = {2007},
	pages = {150201},
}

@article{sartor_mean-field_2021,
	title = {Mean-{Field} {Predictions} of {Scaling} {Prefactors} {Match} {Low}-{Dimensional} {Jammed} {Packings}},
	volume = {126},
	url = {https://link.aps.org/doi/10.1103/PhysRevLett.126.048001},
	doi = {10.1103/PhysRevLett.126.048001},
	number = {4},
	urldate = {2026-07-31},
	journal = {Physical Review Letters},
	publisher = {American Physical Society},
	author = {Sartor, James D. and Ridout, Sean A. and Corwin, Eric I.},
	month = jan,
	year = {2021},
	pages = {048001},
}

@article{smirnov_table_1948,
	title = {Table for {Estimating} the {Goodness} of {Fit} of {Empirical} {Distributions}},
	volume = {19},
	issn = {0003-4851},
	url = {https://www.jstor.org/stable/2236278},
	number = {2},
	urldate = {2026-07-27},
	journal = {The Annals of Mathematical Statistics},
	publisher = {Institute of Mathematical Statistics},
	author = {Smirnov, N.},
	year = {1948},
	pages = {279--281},
}

@incollection{plato_n_2005,
	address = {Amsterdam},
	title = {A.{N}. {Kolmogorov}, \textit{{Grundbegriffe} der wahrscheinlichkeitsrechnung} (1933)},
	isbn = {978-0-444-50871-3},
	url = {https://www.sciencedirect.com/science/article/pii/B978044450871350156X},
	doi = {10.1016/B978-044450871-3/50156-X},
	urldate = {2026-07-27},
	booktitle = {Landmark {Writings} in {Western} {Mathematics} 1640-1940},
	publisher = {Elsevier Science},
	author = {Plato, Jan von},
	editor = {Grattan-Guinness, I. and Cooke, Roger and Corry, Leo and Crépel, Pierre and Guicciardini, Niccolo},
	month = jan,
	year = {2005},
	pages = {960--969},
}

@article{fyodorov_hessian_2018,
	title = {Hessian spectrum at the global minimum of high-dimensional random landscapes},
	volume = {51},
	issn = {1751-8121},
	url = {https://doi.org/10.1088/1751-8121/aae74f},
	doi = {10.1088/1751-8121/aae74f},
	language = {en},
	number = {47},
	urldate = {2026-07-27},
	journal = {Journal of Physics A: Mathematical and Theoretical},
	publisher = {IOP Publishing},
	author = {Fyodorov, Yan V and Le Doussal, Pierre},
	month = oct,
	year = {2018},
	pages = {474002},
}

@misc{auffinger_random_2010,
	title = {Random {Matrices} and complexity of {Spin} {Glasses}},
	url = {https://arxiv.org/abs/1003.1129v2},
	language = {en},
	urldate = {2026-07-27},
	journal = {arXiv.org},
	author = {Auffinger, A. and Arous, G. Ben and Cerny, J.},
	month = mar,
	year = {2010},
}

@article{xu_hessian_2025,
	title = {Hessian spectrum at the global minimum of the spherical pure-like mixed p-spin glasses},
	volume = {66},
	issn = {0022-2488},
	url = {https://pubs.aip.org/aip/jmp/article/66/9/093301/3361659/Hessian-spectrum-at-the-global-minimum-of-the},
	doi = {10.1063/5.0251208},
	language = {en},
	number = {9},
	urldate = {2026-07-27},
	journal = {Journal of Mathematical Physics},
	publisher = {AIP Publishing},
	author = {Xu, Hao and Yang, Haoran},
	month = sep,
	year = {2025},
}

@article{parisi_mean-field_2010,
	title = {Mean-field theory of hard sphere glasses and jamming},
	volume = {82},
	url = {https://link.aps.org/doi/10.1103/RevModPhys.82.789},
	doi = {10.1103/RevModPhys.82.789},
	number = {1},
	urldate = {2026-07-25},
	journal = {Reviews of Modern Physics},
	publisher = {American Physical Society},
	author = {Parisi, Giorgio and Zamponi, Francesco},
	month = mar,
	year = {2010},
	pages = {789--845},
}

@article{crisanti_sphericalp-spin_1992,
	title = {The sphericalp-spin interaction spin glass model: the statics},
	volume = {87},
	issn = {1431-584X},
	shorttitle = {The sphericalp-spin interaction spin glass model},
	url = {https://doi.org/10.1007/BF01309287},
	doi = {10.1007/BF01309287},
	language = {en},
	number = {3},
	urldate = {2026-07-25},
	journal = {Zeitschrift für Physik B Condensed Matter},
	author = {Crisanti, A. and Sommers, H. -J.},
	month = oct,
	year = {1992},
	pages = {341--354},
}

@article{derrida_random-energy_1981,
	title = {Random-energy model: {An} exactly solvable model of disordered systems},
	volume = {24},
	shorttitle = {Random-energy model},
	url = {https://link.aps.org/doi/10.1103/PhysRevB.24.2613},
	doi = {10.1103/PhysRevB.24.2613},
	number = {5},
	urldate = {2026-07-25},
	journal = {Physical Review B},
	publisher = {American Physical Society},
	author = {Derrida, Bernard},
	month = sep,
	year = {1981},
	pages = {2613--2626},
}

@article{derrida_random-energy_1980,
	title = {Random-{Energy} {Model}: {Limit} of a {Family} of {Disordered} {Models}},
	volume = {45},
	shorttitle = {Random-{Energy} {Model}},
	url = {https://link.aps.org/doi/10.1103/PhysRevLett.45.79},
	doi = {10.1103/PhysRevLett.45.79},
	number = {2},
	urldate = {2026-07-25},
	journal = {Physical Review Letters},
	publisher = {American Physical Society},
	author = {Derrida, B.},
	month = jul,
	year = {1980},
	pages = {79--82},
}

@article{nishikawa_collective_2022,
	title = {Collective dynamics in a glass-former with {Mari}-{Kurchan} interactions},
	volume = {156},
	issn = {0021-9606, 1089-7690},
	url = {http://arxiv.org/abs/2204.05130},
	doi = {10.1063/5.0096356},
	language = {en},
	number = {24},
	urldate = {2026-07-07},
	journal = {The Journal of Chemical Physics},
	author = {Nishikawa, Yoshihiko and Ikeda, Atsushi and Berthier, Ludovic},
	month = jun,
	year = {2022},
	note = {arXiv:2204.05130 [cond-mat.dis-nn]},
	pages = {244503},
}

@article{mari_dynamical_2011,
	title = {Dynamical transition of glasses: {From} exact to approximate},
	volume = {135},
	issn = {0021-9606},
	shorttitle = {Dynamical transition of glasses},
	url = {https://doi.org/10.1063/1.3626802},
	doi = {10.1063/1.3626802},
	number = {12},
	urldate = {2026-07-07},
	journal = {The Journal of Chemical Physics},
	author = {Mari, Romain and Kurchan, Jorge},
	month = sep,
	year = {2011},
	pages = {124504},
}

@article{mari_jamming_2009,
	title = {Jamming versus {Glass} {Transitions}},
	volume = {103},
	url = {https://link.aps.org/doi/10.1103/PhysRevLett.103.025701},
	doi = {10.1103/PhysRevLett.103.025701},
	number = {2},
	urldate = {2026-07-07},
	journal = {Physical Review Letters},
	publisher = {American Physical Society},
	author = {Mari, Romain and Krzakala, Florent and Kurchan, Jorge},
	month = jul,
	year = {2009},
	pages = {025701},
}

@article{folena_rethinking_2020,
	title = {Rethinking {Mean}-{Field} {Glassy} {Dynamics} and {Its} {Relation} with the {Energy} {Landscape}: {The} {Surprising} {Case} of the {Spherical} {Mixed} \$p\$-{Spin} {Model}},
	volume = {10},
	shorttitle = {Rethinking {Mean}-{Field} {Glassy} {Dynamics} and {Its} {Relation} with the {Energy} {Landscape}},
	url = {https://link.aps.org/doi/10.1103/PhysRevX.10.031045},
	doi = {10.1103/PhysRevX.10.031045},
	number = {3},
	urldate = {2026-04-28},
	journal = {Physical Review X},
	publisher = {American Physical Society},
	author = {Folena, Giampaolo and Franz, Silvio and Ricci-Tersenghi, Federico},
	month = aug,
	year = {2020},
	pages = {031045},
}

@article{bautista_numerically_2026,
	title = {Numerically discovered inherent states are always protocol dependent in jammed packings},
	volume = {113},
	url = {https://link.aps.org/doi/10.1103/fg38-mtlf},
	doi = {10.1103/fg38-mtlf},
	number = {4},
	urldate = {2026-04-28},
	journal = {Physical Review E},
	publisher = {American Physical Society},
	author = {Bautista, Eddie and Corwin, Eric I.},
	month = apr,
	year = {2026},
	pages = {045408},
}

@article{charbonneau_numerical_2015,
	title = {Numerical detection of the {Gardner} transition in a mean-field glass former},
	volume = {92},
	url = {https://link.aps.org/doi/10.1103/PhysRevE.92.012316},
	doi = {10.1103/PhysRevE.92.012316},
	number = {1},
	urldate = {2025-12-08},
	journal = {Physical Review E},
	publisher = {American Physical Society},
	author = {Charbonneau, Patrick and Jin, Yuliang and Parisi, Giorgio and Rainone, Corrado and Seoane, Beatriz and Zamponi, Francesco},
	month = jul,
	year = {2015},
	pages = {012316},
}

@article{charbonneau_jamming_2015,
	title = {Jamming {Criticality} {Revealed} by {Removing} {Localized} {Buckling} {Excitations}},
	volume = {114},
	url = {https://link.aps.org/doi/10.1103/PhysRevLett.114.125504},
	doi = {10.1103/PhysRevLett.114.125504},
	number = {12},
	urldate = {2025-12-08},
	journal = {Physical Review Letters},
	publisher = {American Physical Society},
	author = {Charbonneau, Patrick and Corwin, Eric I. and Parisi, Giorgio and Zamponi, Francesco},
	month = mar,
	year = {2015},
	pages = {125504},
}

@article{bouchaud_weak_1992,
	title = {Weak ergodicity breaking and aging in disordered systems},
	volume = {2},
	issn = {1155-4304, 1286-4862},
	url = {http://www.edpsciences.org/10.1051/jp1:1992238},
	doi = {10.1051/jp1:1992238},
	language = {en},
	number = {9},
	urldate = {2025-12-08},
	journal = {Journal de Physique I},
	author = {Bouchaud, J. P.},
	month = sep,
	year = {1992},
	pages = {1705--1713},
}

@article{sibani_anomalous_1987,
	title = {Anomalous diffusion and low-temperature spin-glass susceptibility},
	volume = {35},
	url = {https://link.aps.org/doi/10.1103/PhysRevB.35.8572},
	doi = {10.1103/PhysRevB.35.8572},
	number = {16},
	urldate = {2025-12-08},
	journal = {Physical Review B},
	publisher = {American Physical Society},
	author = {Sibani, Paolo},
	month = jun,
	year = {1987},
	pages = {8572--8578},
}

@article{fisher_equilibrium_1988,
	title = {Equilibrium behavior of the spin-glass ordered phase},
	volume = {38},
	url = {https://link.aps.org/doi/10.1103/PhysRevB.38.386},
	doi = {10.1103/PhysRevB.38.386},
	number = {1},
	urldate = {2025-12-08},
	journal = {Physical Review B},
	publisher = {American Physical Society},
	author = {Fisher, Daniel S. and Huse, David A.},
	month = jul,
	year = {1988},
	pages = {386--411},
}

@misc{bautista_numerically_2025,
	title = {Numerically {Discovered} {Inherent} {States} are {Always} {Protocol} {Dependent} in {Jammed} {Packings}},
	url = {http://arxiv.org/abs/2508.09284},
	doi = {10.48550/arXiv.2508.09284},
	urldate = {2025-10-06},
	publisher = {arXiv},
	author = {Bautista, Eddie and Corwin, Eric I.},
	month = sep,
	year = {2025},
	note = {arXiv:2508.09284 [cond-mat]},
}

@article{cavagna_stationary_1998,
	title = {Stationary points of the {Thouless}-{Anderson}-{Palmer} free energy},
	volume = {57},
	url = {https://link.aps.org/doi/10.1103/PhysRevB.57.11251},
	doi = {10.1103/PhysRevB.57.11251},
	number = {18},
	urldate = {2025-08-25},
	journal = {Physical Review B},
	publisher = {American Physical Society},
	author = {Cavagna, Andrea and Giardina, Irene and Parisi, Giorgio},
	month = may,
	year = {1998},
	pages = {11251--11257},
}

@article{kurchan_phase_1996,
	title = {Phase space geometry and slow dynamics},
	volume = {29},
	issn = {0305-4470, 1361-6447},
	url = {http://arxiv.org/abs/cond-mat/9510079},
	doi = {10.1088/0305-4470/29/9/009},
	language = {en},
	number = {9},
	urldate = {2025-08-16},
	journal = {Journal of Physics A: Mathematical and General},
	author = {Kurchan, Jorge and Laloux, Laurent},
	month = may,
	year = {1996},
	note = {arXiv:cond-mat/9510079},
	pages = {1929--1948},
}

@article{dennis_jamming_2020,
	title = {Jamming {Energy} {Landscape} is {Hierarchical} and {Ultrametric}},
	volume = {124},
	issn = {0031-9007, 1079-7114},
	url = {https://link.aps.org/doi/10.1103/PhysRevLett.124.078002},
	doi = {10.1103/PhysRevLett.124.078002},
	language = {en},
	number = {7},
	urldate = {2023-05-17},
	journal = {Physical Review Letters},
	author = {Dennis, R.C. and Corwin, E.I.},
	month = feb,
	year = {2020},
	pages = {078002},
}

@article{urbani_shear_2017,
	title = {Shear {Yielding} and {Shear} {Jamming} of {Dense} {Hard} {Sphere} {Glasses}},
	volume = {118},
	url = {https://link.aps.org/doi/10.1103/PhysRevLett.118.038001},
	doi = {10.1103/PhysRevLett.118.038001},
	number = {3},
	urldate = {2025-07-31},
	journal = {Physical Review Letters},
	publisher = {American Physical Society},
	author = {Urbani, Pierfrancesco and Zamponi, Francesco},
	month = jan,
	year = {2017},
	pages = {038001},
}

@article{rainone_following_2015,
	title = {Following the {Evolution} of {Hard} {Sphere} {Glasses} in {Infinite} {Dimensions} under {External} {Perturbations}: {Compression} and {Shear} {Strain}},
	volume = {114},
	shorttitle = {Following the {Evolution} of {Hard} {Sphere} {Glasses} in {Infinite} {Dimensions} under {External} {Perturbations}},
	url = {https://link.aps.org/doi/10.1103/PhysRevLett.114.015701},
	doi = {10.1103/PhysRevLett.114.015701},
	number = {1},
	urldate = {2025-07-31},
	journal = {Physical Review Letters},
	publisher = {American Physical Society},
	author = {Rainone, Corrado and Urbani, Pierfrancesco and Yoshino, Hajime and Zamponi, Francesco},
	month = jan,
	year = {2015},
	pages = {015701},
}

@article{ashwin_calculations_2012,
	title = {Calculations of the {Structure} of {Basin} {Volumes} for {Mechanically} {Stable} {Packings}},
	volume = {85},
	issn = {1539-3755, 1550-2376},
	url = {http://arxiv.org/abs/1112.4234},
	doi = {10.1103/PhysRevE.85.061307},
	number = {6},
	urldate = {2025-07-31},
	journal = {Physical Review E},
	author = {Ashwin, S. S. and Blawzdziewicz, J. and O'Hern, C. S. and Shattuck, M. D.},
	month = jun,
	year = {2012},
	note = {arXiv:1112.4234 [cond-mat]},
	pages = {061307},
}

@misc{arceri_jamming_2022,
	title = {The {Jamming} {Transition} and the {Marginally} {Stable} {Solid}},
	url = {http://arxiv.org/abs/2209.02829},
	doi = {10.48550/arXiv.2209.02829},
	urldate = {2025-07-31},
	publisher = {arXiv},
	author = {Arceri, Francesco and Corwin, Eric I. and O'Hern, Corey S.},
	month = sep,
	year = {2022},
	note = {arXiv:2209.02829 [cond-mat]},
}

@article{martiniani_numerical_2017,
	title = {Numerical test of the {Edwards} conjecture shows that all packings are equally probable at jamming},
	volume = {13},
	copyright = {2017 Springer Nature Limited},
	issn = {1745-2481},
	url = {https://www.nature.com/articles/nphys4168},
	doi = {10.1038/nphys4168},
	language = {en},
	number = {9},
	urldate = {2025-07-31},
	journal = {Nature Physics},
	publisher = {Nature Publishing Group},
	author = {Martiniani, Stefano and Schrenk, K. Julian and Ramola, Kabir and Chakraborty, Bulbul and Frenkel, Daan},
	month = sep,
	year = {2017},
	pages = {848--851},
}

@article{speedy_random_1998,
	title = {Random jammed packings of hard discs and spheres},
	volume = {10},
	issn = {0953-8984},
	url = {https://dx.doi.org/10.1088/0953-8984/10/19/006},
	doi = {10.1088/0953-8984/10/19/006},
	language = {en},
	number = {19},
	urldate = {2025-07-31},
	journal = {Journal of Physics: Condensed Matter},
	author = {Speedy, Robin J.},
	month = may,
	year = {1998},
	pages = {4185},
}

@article{charbonneau_jamming_2023,
	title = {Jamming, relaxation, and memory in a minimally structured glass former},
	volume = {108},
	url = {https://link.aps.org/doi/10.1103/PhysRevE.108.054102},
	doi = {10.1103/PhysRevE.108.054102},
	number = {5},
	urldate = {2025-07-30},
	journal = {Physical Review E},
	publisher = {American Physical Society},
	author = {Charbonneau, Patrick and Morse, Peter K.},
	month = nov,
	year = {2023},
	pages = {054102},
}

@article{charbonneau_fractal_2014,
	title = {Fractal free energy landscapes in structural glasses},
	volume = {5},
	copyright = {2014 Springer Nature Limited},
	issn = {2041-1723},
	url = {https://www.nature.com/articles/ncomms4725},
	doi = {10.1038/ncomms4725},
	language = {en},
	number = {1},
	urldate = {2025-07-14},
	journal = {Nature Communications},
	publisher = {Nature Publishing Group},
	author = {Charbonneau, Patrick and Kurchan, Jorge and Parisi, Giorgio and Urbani, Pierfrancesco and Zamponi, Francesco},
	month = apr,
	year = {2014},
	pages = {3725},
}

@article{castellani_spin-glass_2005,
	title = {Spin-{Glass} {Theory} for {Pedestrians}},
	volume = {2005},
	issn = {1742-5468},
	url = {http://arxiv.org/abs/cond-mat/0505032},
	doi = {10.1088/1742-5468/2005/05/P05012},
	number = {05},
	urldate = {2025-07-14},
	journal = {Journal of Statistical Mechanics: Theory and Experiment},
	author = {Castellani, Tommaso and Cavagna, Andrea},
	month = may,
	year = {2005},
	note = {arXiv:cond-mat/0505032},
	pages = {P05012},
}

@article{sciortino_potential_2005,
	title = {Potential energy landscape description of supercooled liquids and glasses},
	volume = {2005},
	issn = {1742-5468},
	url = {https://dx.doi.org/10.1088/1742-5468/2005/05/P05015},
	doi = {10.1088/1742-5468/2005/05/P05015},
	language = {en},
	number = {05},
	urldate = {2025-07-11},
	journal = {Journal of Statistical Mechanics: Theory and Experiment},
	author = {Sciortino, Francesco},
	month = may,
	year = {2005},
	pages = {P05015},
}

@article{martiniani_when_2023,
	title = {When you can’t count, sample! {Computable} entropies beyond equilibrium from basin volumes},
	volume = {15},
	copyright = {Copyright (c) 2023 Stefano Martiniani, Mathias Casiulis},
	issn = {1852-4249},
	url = {https://www.papersinphysics.org/papersinphysics/article/view/837},
	doi = {10.4279/pip.150001},
	language = {en},
	urldate = {2023-05-17},
	journal = {Papers in Physics},
	author = {Martiniani, Stefano and Casiulis, Mathias},
	month = feb,
	year = {2023},
	pages = {150001--150001},
}

@article{hwang_understanding_2016,
	title = {Understanding soft glassy materials using an energy landscape approach},
	volume = {15},
	copyright = {2016 Nature Publishing Group},
	issn = {1476-4660},
	url = {https://www.nature.com/articles/nmat4663},
	doi = {10.1038/nmat4663},
	language = {en},
	number = {9},
	urldate = {2023-05-17},
	journal = {Nature Materials},
	publisher = {Nature Publishing Group},
	author = {Hwang, Hyun Joo and Riggleman, Robert A. and Crocker, John C.},
	month = sep,
	year = {2016},
	note = {Number: 9},
	pages = {1031--1036},
}

@misc{suryadevara_basins_2025,
	title = {The {Basins} of {Attraction} of {Soft} {Sphere} {Packings} {Are} {Not} {Fractal}},
	url = {http://arxiv.org/abs/2409.12113},
	doi = {10.48550/arXiv.2409.12113},
	urldate = {2025-07-08},
	publisher = {arXiv},
	author = {Suryadevara, Praharsh and Casiulis, Mathias and Martiniani, Stefano},
	month = jul,
	year = {2025},
	note = {arXiv:2409.12113 [cond-mat]},
}

@article{cauchy_methode_nodate,
	title = {Me´thode ge´ne´rale pour la re´solution des syste`mes d’e´quations simultane´es},
	language = {en},
	author = {Cauchy, M Augustine},
}

@article{charbonneau_glass_2017,
	title = {Glass and {Jamming} {Transitions}: {From} {Exact} {Results} to {Finite}-{Dimensional} {Descriptions}},
	volume = {8},
	shorttitle = {Glass and {Jamming} {Transitions}},
	url = {https://doi.org/10.1146/annurev-conmatphys-031016-025334},
	doi = {10.1146/annurev-conmatphys-031016-025334},
	number = {1},
	urldate = {2023-06-08},
	journal = {Annual Review of Condensed Matter Physics},
	author = {Charbonneau, Patrick and Kurchan, Jorge and Parisi, Giorgio and Urbani, Pierfrancesco and Zamponi, Francesco},
	year = {2017},
	note = {\_eprint: https://doi.org/10.1146/annurev-conmatphys-031016-025334},
	pages = {265--288},
}

@article{thirumalaiswamy_exploring_2022,
	title = {Exploring canyons in glassy energy landscapes using metadynamics},
	volume = {119},
	issn = {0027-8424, 1091-6490},
	url = {http://arxiv.org/abs/2204.00587},
	doi = {10.1073/pnas.2210535119},
	number = {43},
	urldate = {2023-05-17},
	journal = {Proceedings of the National Academy of Sciences},
	author = {Thirumalaiswamy, Amruthesh and Riggleman, Robert A. and Crocker, John C.},
	month = oct,
	year = {2022},
	note = {arXiv:2204.00587 [cond-mat]},
	pages = {e2210535119},
}

@article{charbonneau_exact_2014,
	title = {Exact theory of dense amorphous hard spheres in high dimension. {III}. {The} full {RSB} solution},
	volume = {2014},
	issn = {1742-5468},
	url = {http://arxiv.org/abs/1310.2549},
	doi = {10.1088/1742-5468/2014/10/P10009},
	number = {10},
	urldate = {2023-05-17},
	journal = {Journal of Statistical Mechanics: Theory and Experiment},
	author = {Charbonneau, Patrick and Kurchan, Jorge and Parisi, Giorgio and Urbani, Pierfrancesco and Zamponi, Francesco},
	month = oct,
	year = {2014},
	note = {arXiv:1310.2549 [cond-mat]},
	pages = {P10009},
}

@article{morse_echoes_2017,
	title = {Echoes of the {Glass} {Transition} in {Athermal} {Soft} {Spheres}},
	volume = {119},
	url = {https://link.aps.org/doi/10.1103/PhysRevLett.119.118003},
	doi = {10.1103/PhysRevLett.119.118003},
	number = {11},
	urldate = {2025-07-09},
	journal = {Physical Review Letters},
	publisher = {American Physical Society},
	author = {Morse, Peter K. and Corwin, Eric I.},
	month = sep,
	year = {2017},
	pages = {118003},
}

@article{artiaco_exploratory_2020,
	title = {An exploratory study of the glassy landscape near jamming},
	volume = {101},
	issn = {2470-0045, 2470-0053},
	url = {http://arxiv.org/abs/1908.06127},
	doi = {10.1103/PhysRevE.101.052605},
	language = {en},
	number = {5},
	urldate = {2023-06-07},
	journal = {Physical Review E},
	author = {Artiaco, Claudia and Baldan, Paolo and Parisi, Giorgio},
	month = may,
	year = {2020},
	note = {arXiv:1908.06127 [cond-mat]},
	pages = {052605},
}

@article{charbonneau_universal_2016,
	title = {Universal {Non}-{Debye} {Scaling} in the {Density} of {States} of {Amorphous} {Solids}},
	volume = {117},
	url = {https://link.aps.org/doi/10.1103/PhysRevLett.117.045503},
	doi = {10.1103/PhysRevLett.117.045503},
	number = {4},
	urldate = {2025-07-09},
	journal = {Physical Review Letters},
	publisher = {American Physical Society},
	author = {Charbonneau, Patrick and Corwin, Eric I. and Parisi, Giorgio and Poncet, Alexis and Zamponi, Francesco},
	month = jul,
	year = {2016},
	pages = {045503},
}

@article{morse_geometric_2014,
	title = {Geometric {Signatures} of {Jamming} in the {Mechanical} {Vacuum}},
	volume = {112},
	url = {https://link.aps.org/doi/10.1103/PhysRevLett.112.115701},
	doi = {10.1103/PhysRevLett.112.115701},
	number = {11},
	urldate = {2025-07-09},
	journal = {Physical Review Letters},
	publisher = {American Physical Society},
	author = {Morse, Peter K. and Corwin, Eric I.},
	month = mar,
	year = {2014},
	pages = {115701},
}

@article{gardner_spin_1985,
	title = {Spin glasses with p-spin interactions},
	volume = {257},
	issn = {0550-3213},
	url = {https://www.sciencedirect.com/science/article/pii/0550321385903748},
	doi = {10.1016/0550-3213(85)90374-8},
	language = {en},
	urldate = {2023-05-17},
	journal = {Nuclear Physics B},
	author = {Gardner, E.},
	month = jan,
	year = {1985},
	pages = {747--765},
}

\end{document}